\documentclass{article}
\usepackage{graphicx}
\usepackage{url}

\title{An $O(k \log{n})$ algorithm for prefix based ranked
  autocomplete}

\author{Dhruv Matani}

\begin{document}
\date{September 2, 2011}

\maketitle

\begin{abstract}
Many search engines such as Google, Bing \& Yahoo! show search
suggestions when users enter search phrases on their interfaces. These
suggestions are meant to assist the user in finding what she wants
quickly and also suggesting common searches that may result in finding
information that is more relevant. It also serves the purpose of
helping the user if she is not sure of what to search for, but has a
vague idea of what it is that she wants. We present an algorithm that
takes time proportional to $O(k \log{n})$, and $O(n)$ extra space for
providing the user with the top $k$ ranked suggestions out of a corpus
of $n$ possible suggestions based on the prefix of the query that she
has entered so far.

\end{abstract}

\clearpage

\tableofcontents

\clearpage


\section{What is prefix based ranked phrase auto-complete?}

Given a set of $n$ strings, each having a certain
\textit{weight}, the problem of finding the set of $k$ heaviest
strings each of which have the prefix $q$ is the problem of
prefix based ranked phrase auto-complete.

\section{Where is phrase auto-complete used?}

Many search engines such as Google, Bing \& Yahoo! show search
suggestions when users enter search phrases on their interfaces. These
suggestions are meant to assist the user in finding what she wants
quickly and also suggesting common searches that may result in finding
information that is more relevant. It also serves the purpose of
helping the user if she is not sure of what to search for, but has a
vague idea of what it is that she wants.

IMDB uses search suggestions to ensure faster search results since
movie titles are mostly unique after the first few characters of
typing. Users that find it difficult to type or users on mobile
devices with constrained input methods can get to their results much
faster because of auto-complete.

\section{What are the problems that phrase auto-complete solves?}

Many a times, the user can think of many keywords or phrases that can
be used to describe the concept or idea that is being searched
for. Such descriptions generally use similar meaning words. Users
don't know in advance what vocabulary most of the literature or prior
work in that field uses. Hence, they have to painstakingly try each
combination of keywords till they fine what they are looking for.

For example, when a user is searching for \textit{auto complete}, it
could actually be referred to by articles online as either
\textit{autocomplete}, or \textit{auto suggest}, or \textit{search
  suggestions}, or \textit{find as you type}. The user would have to
try all of them before settling on the one that returns the most
relevant results. If most of the alternatives, 3 in this case, have
the same prefix, then the user could just start typing, and the system
could suggest possible completions as new characters are entered. This
greatly reduces the trial \& error that the user has to
perform. Furthermore, many users aren't even aware of entering
different search terms. With a high probability, these users will not
find what they are looking for.

\section{Why does auto-complete need to be fast (responsive)?}




The average typing speed (while composing text) is 19 words per
minute\cite{wikiWPM}. The average word length in the English language
is about 9 characters\cite{tb48soj2}. Combining the two, we notice
that the average time for typing a single character is
\textit{351ms}. Accounting for a network round-trip time of
\textit{200ms}\cite{SMXX,rfc619}, and client processing time of
\textit{100ms}, we have only about \textit{50ms} left to do the
processing at our end. The auto complete application needs to return a
list of suggestions within \textit{50ms} for it to be useful to the
user. It needs to do this for potentially every keystroke typed in by
the user (there are optimizations we can do at the user interface
layer, but we won't discuss them here).

\section{Problem Statement}

Given a list of $n$ phrases, along with a weight for each phrase and a
query prefix $q$, determine the $k$ heaviest phrases that have $q$ as
their prefix. We assume the average phrase length to be constant and
shall not account for it in the complexity calculations.

\section{Existing approaches}

We discuss three existing approaches used by various applications to
provide search suggestion for a \textit{find as you type} experience
on their web pages.

\subsection{Naive approach}

The naive approach involves pre-processing the input ($n$ phrases) by
sorting them in lexicographical order. Binary Search is then used to
locate the beginning and end of the candidate list of phrases. Each
phrase in this candidate list $cl$ has $q$ as its prefix.

Now, all the terms in the candidate list $cl$ are sorted in
non-increasing order by weight and the top $k$ are selected for
projection.

This requires $O(log{n})$ for the binary search and $O(cl \log{cl})$
for sorting the candidate list. This approach also requires extra
space proportional to $O(cl)$ for storing the candidate list $cl$
before sorting. The total complexity of this approach is $O(\log{n}) +
O(cl \log{cl})$.

We can see that if the query prefix $q$ is short or matches many
phrases, then the candidate list $cl$ will be large and the latter
factor will dominate the complexity. We ideally want an algorithm that
is not input-sensitive.

A slight variation on the above technique would be to notice that $k$
is generally quite small (around 16), so fetching the top 16
candidates from the candidate list of size $cl$ will cost only
$O(k.cl)$ which is faster than what we have before. This brings our
runtime down to $O(\log{n}) + O(k.cl)$. However, it is still
input-sensitive and that is undesirable.

\subsection{Space intensive approach}


We could maintain a lexicographically sorted list of all possible
phrase prefixes, with all the entries with the same prefix sorted by
weight. This implies sorting by two keys, namely \textit{prefix,
  weight}. A lookup is as simple as performing a binary search to
locate the first occurrence of the query $q$ and reading off the next
$k$ entries, as long as they have the same value as the query $q$,
since they are already sorted by weight.

This technique uses extra space that is proportional to
$O(string\ length)$ for each phrase in the initial set of $n$
phrases. This means that extra space to the order of
$O(n.string\ length / 2)$, which roughly translates to $15n$ in our
case, is needed. We can not ignore the constant factor here since it
is very significant. For an initial corpus of $5GB$ (17 million
phrases of length 30 characters each), we would land up using extra
space proportional to $75GB$ in our case.

The runtime complexity for the pre-processing step is $O(15n)$ and the
runtime complexity for querying is $O(k \log{n})$.

This approach is attractive for small data sets, but starts getting
very costly in terms of memory requirements for larger data sets.

This approach is used by the \textit{redis} data structure store for
providing auto-complete facilities\cite{redisAC}.

\subsection{Ternary search trees}

We can optimize the previous approach for space by using a
\textit{trie} data structure instead of duplicating each prefix of a
phrase every time we encounter it. This gives rise to the
\textit{ternary tree} data structure. We need to decide in advance
what the maximum value of $k$ is that we would like to support since
this method involves pre-computing a list of the $k$ heaviest
completions for a given prefix and storing them at that node. Since
the actual completions can be stored elsewhere, we incur a penalty
for storing $k$ pointers to phrases at every ternary tree node, and
not a penalty for the $k$ complete phrases. We also incur a penalty
for unused trie-nodes at every step (in the implementation that trades
space for speed).

We can compute the extra space required by assuming that we would need
an average of 10 pointers (and not 16) at every ternary tree
node. Since the ternary tree structure is highly input dependent, we
can't really perform an estimation here. We abandon this approach
since it is highly input dependant and can degenerate to requiring a
lot of space.

\subsection{The TASTIER approach}

Researchers from Tsinghua University and The University of California,
Irvine have implemented a system for predicting user input and
performing an auto-complete based on partial input from the
user\cite{tastier2009}. Their technique differs from other (and the one presented here)
in that it treats the query as a set of keywords, and support
prediction of any combination of keywords that may not be close to
each other in the data. Other techniques (and the one presented here)
do prediction using sequential data to construct their model,
therefore they can only predict a word or phrase that matches
sequentially with underlying data (e.g., phrases or sentences).

However, their technique relies on:
\begin{itemize}
\item Maintaining the previous result set for a past query from the
  user. This increases memory costs on the server and leads to
  increased memory pressure when serving many concurrent users
\item Performing a union of potentially many candidate lists (which is
  time consuming)
\end{itemize}

On a corpus of 1 million entries, their system is able to answer $1$
query per $20$ ms. The result set size here is $10$. This means that
their system can handle at best $50$ qps, which is too low for
search-engine traffic.

\section{Proposed algorithm}

We present an algorithm that takes time proportional to $O(k
\log{n})$, and $O(n)$ extra space for providing the user with the top
$k$ ranked suggestions out of a corpus of $n$ possible suggestions
based on the prefix of the query that she has entered so far.

\begin{figure}[h!]
\centering 
\includegraphics[width=12cm]{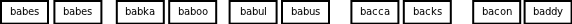}
\caption{The sorted array of all completion phrases}
\label{fig:sortedwords}
\end{figure}

This algorithm requires us to maintain the set of phrases in a sorted
array (Figure ~\ref{fig:sortedwords}) so that we can \textit{binary
  search} over it using our query prefix \textit{q}.

\begin{figure}[h!]
\centering
\includegraphics[width=12cm]{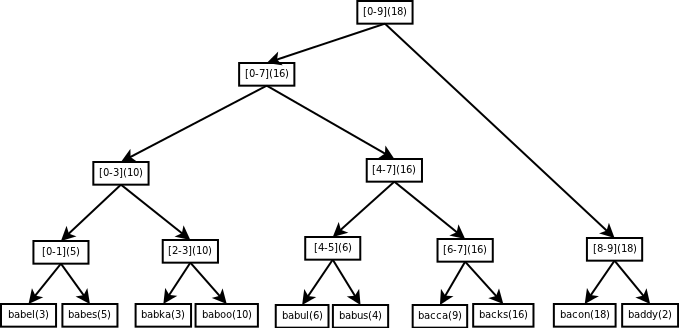}
\caption{The Segment Tree keyed by phrase frequency, shown in
  parenthesis after each word. The internal nodes are of the format
  [range](maximum key in subtree)}
\label{fig:segmenttree}
\end{figure}

We additionally also maintain a \textit{Segment Tree}\cite{segTree}
(Figure ~\ref{fig:segmenttree}) (which is nothing but an
\textit{Interval Tree}) that stores the maximum weight of the interval
at every node. In the referenced figure, each leaf node shows the word
that the node represents, though the word is actually not stored at
that node. Instead, the number in parenthesis is stored in the node,
and it denotes the weight of the phrase that the node is associated
with. Each internal node shows a range (in square brackets) denoting
the lower and upper indexes (both inclusive) in the sorted word array
that this internal node represents. Each internal node stores the
maximum weight of the nodes in the range. This weight is shown in
parenthesis in the diagram.

While querying, we maintain an ephemeral max-heap (or max-priority queue) of
ranges that contain the maximum weighted candidate phrases. The nodes
of this max-heap are keyed by the maximum weight of phrases stored by
the range of indexes that the node represents. For example, if the
node represents range [0--20] and its key is $45$, it means that the
\textit{heaviest} phrase in the range [0--20] has a weight of $45$. We
always start by entering a single element representing the entire
range of interest into the max-heap and then spliting the range
containing the heaviest phrase at every step.

\begin{figure}[h!]
\centering
\includegraphics[width=12cm]{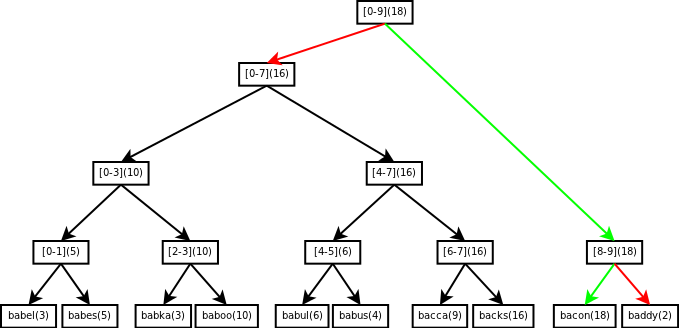}
\caption{Search for the maximum weighted phrase in the range
  [0--9]. The green arrows show the paths taken and the red arrows
  show the paths not taken}
\label{fig:lookup09}
\end{figure}

\begin{figure}[h!]
\centering
\includegraphics[width=12cm]{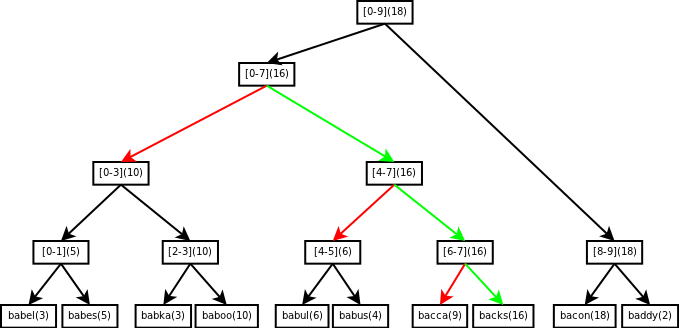}
\caption{Search for the maximum weighted phrase in the range
  [0--7]. The green arrows show the paths taken and the red arrows
  show the paths not taken}
\label{fig:lookup07}
\end{figure}

\begin{figure}[h!]
\centering
\includegraphics[width=12cm]{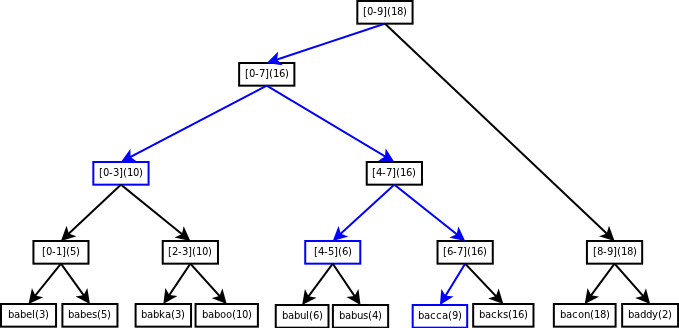}
\caption{Search for the maximum weighted phrase in the range
  [0--6]. The blue arrows show the paths and branches taken. With the
  help of the blue arrows, we are able to trace the complete path of
  the query to the leaf nodes. The blue boses indicate the nodes where
  the query for the maximum weighted node will be restarted.}
\label{fig:lookup06}
\end{figure}

\begin{figure}[h!]
\centering
\includegraphics[width=12cm]{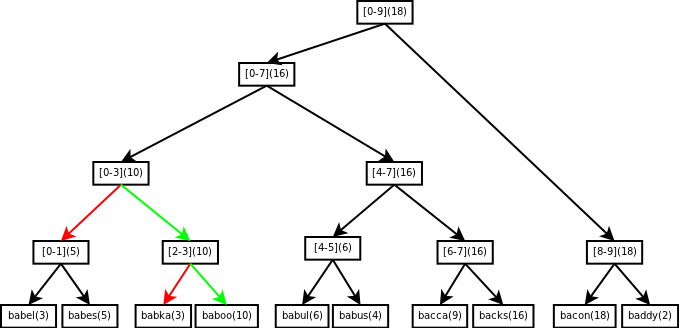}
\caption{Search for the maximum weighted phrase in the range
  [0--3]. The green arrows show the paths taken and the red arrows
  show the paths not taken.}
\label{fig:lookup03}
\end{figure}

For the example shown in Figure ~\ref{fig:lookup09}, we show the first
step in the series of operations performed on the \textit{Segment
  Tree} while searching for the top $4$ phrases that begin with the
string \textit{b} in the set shown in Figure
~\ref{fig:sortedwords}. The search starts by inserting the entire
range of indices that contain the candidate words ([0-9] in this case)
into the max-heap. The top of the max-heap is then popped and the
index of the highest phrase in that range is selected. We see that the
word \textit{bacon} with the weight \textit{18} is selected and a
split at the index \textit{8} gives rise to 2 ranges, [0-7] \&
[9-9]. These are inserted into the max-heap and the procedure is
repeated. Figures ~\ref{fig:lookup07}, ~\ref{fig:lookup06}, \&
~\ref{fig:lookup03} show the action of the algorithm on the sample
data set.

\clearpage

\begin{figure}[h!]
\centering
\includegraphics[width=10cm]{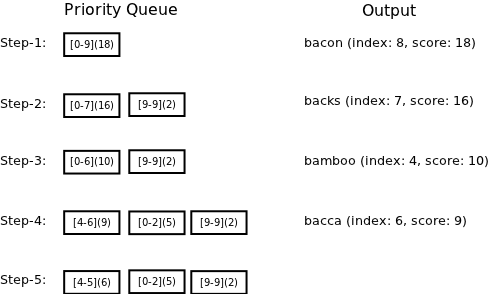}
\caption{}
\label{fig:prioqueue}
\end{figure}

Figure ~\ref{fig:prioqueue} shows the contents of the max-heap at
every step of the process of extracting the top $4$ phrases having the
prefix $b$. The left column indicates the state of the priority queue
(a max heap) at every stage and the right column indicates the output
produced (phrase) at every stage along with the index and weight
(score) of that phrase. The output is generated in non-increasing
order of weight.

\clearpage

\section{Approximate matching for auto-complete}

Researchers from Tsinghua University and The University of California,
Irvine have implemented a system for performing a \textit{fuzzy
  keyword match} between the user input and the indexed
corpus\cite{tastierfuzzy2009}. Their technique relies on creating
multiple keyword candidate lists per keyword entered and intersecting
them to produce the output. They optimize this process by maintaining
cached result sets for previous input by the same user so that they
can just trim these lists when a new character is added.

On a data set with 1 million entries, their technique takes up to
$5ms$ for finding up to $10$ results having prefixes of length $3$
($3$ characters in length). This case is sure to be hit because users
start off by typing at least $2-3$ characters of their query before
hoping for relevant suggestions. In the worst case, this system can
handle a load of $200$ qps. For longer prefixes (i.e. upwards of $6$
characters in length), the query time drops shapply to below $0.1ms$,
which enables them to answer about $10,000$ qps. The responsiveness of
our technique on the other hand is independent of the number of
characters entered by the user and depends only on the corpus
size. Hence, it's responsiveness is much more predictable, which is
desirable in a production environment.

\subsection{Using exact prefix-match auto-complete for approximate match
  auto-complete}

If we pre-process the data and queries in some specific manner, we can
use the exact-match auto-complete implementation to perform
approximate match auto-complete\footnote{This need not be an
  approximate prefix-match auto-complete depending on how we
  pre-process the data}. This can be accomplished by performing one or
more of the following transformations on the phrases in the
corpus as well as the query string. The transformations must be
performed in the same order as given below:

\begin{itemize}

\item Remove all stop-words such as \textit{a, the, have, has, of, etc
  \ldots{}}

\item Remove all punctuation marks, vowels, and white-spaces from the phrase

\item (optionally) Convert all consonants to their numeric equivalents
  according to the transformations mentioned in the \textit{Soundex
    Algorithm}\footnote{US patent 1261167, R. C. Russell, issued
    1918-04-02 \& US patent 1435663, R. C. Russell, issued 1922-11-14}

\item Collapse all repetitive runs of characters (or numbers) into a
  single character (or number). e.g. \textit{rttjdddl} becomes
  \textit{rtjdl}

\end{itemize}

Now, these transformations are stored as the phrase in the
auto-complete index. The original (untransformed) phrase is stored
against the transformed phrase so that once the lookup is done, the
original phrase may be returned.

When we query the auto-complete index, we perform the same transform
on the query string and find those strings that share a prefix with
the query string.

In practice, we can index phrases at every stage of the
transformations and take the union of the result set after querying
each index for the corresponding query string.

\section{Implementation \& Performance Tests}

The idea mentioned above has been implemented in an application called
\textit{lib-face}\footnote{Available for download at
  \url{https://code.google.com/p/lib-face/}}. \textit{lib-face} is
written in the C++ programming language and uses the
\textit{mongoose} web-server\footnote{mongoose is an in-process HTTP
  web-server available for download at
  \url{http://code.google.com/p/mongoose/}} to serve requests. Once
you load the corpus into lib-face, you can query the highest weighted
phrase that has a certain prefix by sending an HTTP request to it.\\
\\
Here are the results of a test run on an Amazon EC2 instance:\\
\\
\begin{tabular}[h]{|l|r|}
  \hline
  Operating System & Ubuntu 10.04 \\
  \hline
  Data Set Size (number of entries) & 14,000,000 \\
  \hline
  Data Set Size (bytes) & 312MiB \\
  \hline
  Result set size & 32\\
  \hline
  Queries Per Second & 6,800 \\
  \hline
  Memory (Resident Size) & 1,540MiB\\
  \hline
\end{tabular}\\
\\
\textit{Notes:}
\begin{enumerate}

\item Not all of the 1,540MiB of resident memory is used since C++'s
  std::vector uses a doubling strategy to grow, which results in
  half the memory being actually unused.

\item There are CPU and network overheads when running an application
  on a virtual machine v/s on real hardware. We think it's best to
  benchmark on a VM rather than real hardware since that seems to be a
  very common deployment paradigm these days. Comparing with
  benchmarks running on real-hardware would be unfair

\item All our benchmarks are done assuming a result set size of $32$
  whereas other groups have assumed a result set size of $10$. This
  will invariably reduce the number of queries per second that we can
  answer.

\end{enumerate}

\section{Conclusion}

The \textit{Segment Tree} method we described above is used
essentially for performing a \textit{range-max query} over a certain
set of phrase weights. There is another method\cite{BC2000} of
performing range-max queries which has a runtime cost of $O(1)$ rather
than $O(\log{n})$ per query. If this method is used, we can reduce the
runtime cost of our method further from $O(k \log{n})$ to $O(k
\log{k})$.

If we use a \textit{suffix array} to represent all the phrases in our
corpus, we can perform a match of the query with a prefix of
\textit{every} suffix of each phrase in the corpus. This makes our
searches even more robus without significantly increasing our
pre-processing or query time. However, we would now need to store an
entry for every starting position of a string in the suffix array in
in our \textit{RQM} data structure. This would increase the size of
the \textit{Segment Tree} to $O(nk)$, where $k$ is the average length
of a phrase in our corpus. This will also have an effect on the query
time, taking it up to $O(\log{nk})$ from $O(\log{n})$.

\clearpage

\bibliographystyle{amsplain}
\bibliography{autocomplete}


\end{document}